\documentclass[aps,prl,twocolumn,showpacs,amsmath,a4paper,amssymb,superscriptaddress,10pt]{revtex4-1}
\usepackage{graphicx}
\usepackage{CJK}
\usepackage{epstopdf}
\usepackage{multirow}

\usepackage{amsmath,amssymb}
\usepackage{color}

\usepackage[usenames,dvipsnames]{xcolor}

\usepackage{natbib}

\def\Wcm2{W/cm$^2$}
\newcommand{\coo}{CO$_2$\,}
\newcommand{\coop}{CO$_2^{+}$\,}
\newcommand{\coopp}{CO$_2^{2+}$\,}

\begin{document}
\begin{CJK*}{UTF8}{gbsn}

\title{Laser-Induced Dissociative Recombination of Carbon Dioxide}

\author{Hongtao Hu}
\author{Seyedreza Larimian}
\author{Sonia Erattupuzha}
\affiliation{Photonics Institute, Technische Universit\"at Wien, A-1040 Vienna, Austria}
\author{Jin Wen}
\email[Corresponding author: ]{wen@uochb.cas.cz}
\affiliation{Institute of Organic Chemistry and Biochemistry, AS CR, 166 10 Praha 6, Czech Republic}
\author{Andrius Baltu\v{s}ka}
\author{Markus Kitzler-Zeiler}
\affiliation{Photonics Institute, Technische Universit\"at Wien, A-1040 Vienna, Austria}
\author{Xinhua Xie (谢新华)}
\email[Corresponding author: ]{xinhua.xie@tuwien.ac.at}
\affiliation{Photonics Institute, Technische Universit\"at Wien, A-1040 Vienna, Austria}
\affiliation{SwissFEL, Paul Scherrer Institute, 5232 Villigen PSI, Switzerland}

\begin{abstract}

We experimentally investigate laser-induced dissociative recombination of CO$_2$ in linearly polarized strong laser fields with coincidence measurements. Our results show laser-induced dissociation processes originate from an electron recombination process after laser-induced double ionization. After double ionization of CO$_2$, one electron is recaptured by the CO$_2^{2+}$ and localized to O$^+$ or CO$^+$ in the following dissociation process.
We found that the probability of electron localization to O$^{+}$ is much higher than that to CO$^+$. Further, our measurements reveal that the recombination probability of the first ionized electron is three times as high as that of the second ionized electron. Our work may trigger further experimental and theoretical studies on involved nuclear and electron dynamics in laser-induced dissociative recombination of molecules and their applications in controlling molecular dissociation with ultrashort laser pulses.

\end{abstract}

\date{\today}




\maketitle
\end{CJK*}

Dissociative recombination refers to the process where a positively charged molecular ion captures a free electron, upon which a (highly) excited molecular complex is formed that subsequently dissociates into fragments \cite{Larsson1997,Seiersen2003}. It is one of the most important processes in plasmas, for instance in the planetary atmospheres, fusion plasmas, and laser physics \cite{Zajfman1996,Fox2014}. Most of experimental studies on dissociative recombination were performed at large facilities, such as heavy-ion storage rings in which the molecular ions are produced by electron-impact ionization \cite{Larsson1997,Seiersen2003}.

In the past decades, femtosecond lasers became a versatile tool to reveal nuclear and electron dynamics of atoms and molecules \cite{Pazourek2015,Atto1_Corkum2007,Posthumus2004,Brabec2000}, due to their ultrashort pulse durations and high peak intensities. When exposed to a strong laser field, molecules can become ionized or excited, which may cause further molecular reactions, such as dissociation and isomerization.
Laser-induced dissociative recombination (LIDR) is one of such reactions, which has attracted the interest of researchers for a decade \cite{Nubb2008}. In the literature on strong-field laser science this process is also referred to dissociative frustrated ionization \cite{Mansch2009}. In this process, electrons are released from a molecule by the strong laser field and afterwards some of them recombine with their parent ion, result in the formation of highly excited dissociative cation. Eventually, the excited molecular cation breaks into fragments.
So far, the research on LIDR mainly focused on homonuclear diatomic molecules, such as H$_2$, D$_2$, and Ar$_2$ \cite{Zhang2017,Wu2011,McKenna2011,Zhang2018}.

In this work, we investigated the LIDR of a triatomic molecule, CO$_2$, in linearly polarized laser fields with a reaction microscope.
As compared to a diatomic, the complexity of a triatomic molecule is significantly higher since many more modes of motion become available \cite{Xie2015}.
Dissociative recombination of CO$_2$ are expected to play an important role in the interaction of CO$_2$ with photons and electrons \cite{Vinci2005}.
Gaining knowledge on dissociative recombination of CO$_2$ is valuable for understanding the concentration of CO$_2$ in planetary atmospheres \cite{Geppert2008}. For example, this process is regarded as a potential source of thermal and non-thermal carbon on Mars \cite{Fox2004}.
Furthermore, the breakage of CO$_2$ can be asymmetric. This allows us to study the electronic localization to different ionic fragments during the dissociation and therewith to extend the knowledge on laser-induced molecular dissociation.



\begin{figure}[htbp]
	\centering
	\includegraphics[width=0.9\columnwidth]{{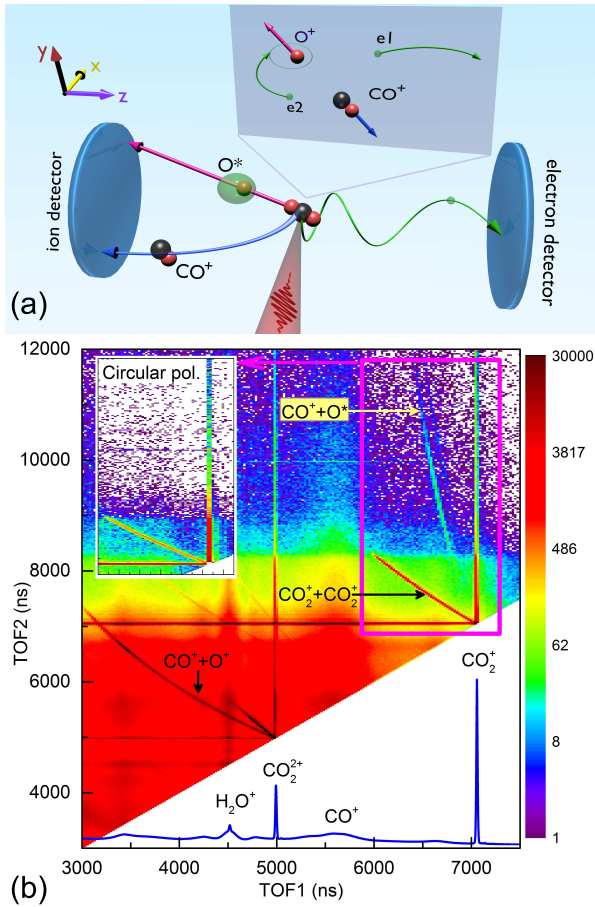}}
	\caption{(a) LIDR of CO$_2$ induced by the intense laser field, and the coincidence detection of resulting particles in a reaction microscope. (b) Measured PIPICO distribution of CO$_2$ in linearly polarized laser field. The blue curve at the bottom of panel (b) is the time-of-flight (TOF) spectrum of the experiment. The insert of panel (b) shows the PIPICO distribution for the measurement with circular polarized laser pulses in the range marked by the pink rectangle.}
	\label{fig1}
\end{figure}

In the experiment, coincidence measurements of three-dimensional momenta of resulting particles are achieved using a reaction microscope \cite{Ullrich2003}.
A schematics of the coincidence measurement of the LIDR process is shown in Fig. \ref{fig1} (a). Charged particles, e.g. CO$^+$ and electrons, are guided to the multihit position- and time-sensitive detectors by a homogeneous dc field of 10.5 V/cm and a uniform magnetic field of 12 G.
Excited neutral particles, e.g. O$^*$, with an initial momentum vector pointing towards the detector are registered with a smaller detection angle as compared to that for charged particles \cite{Nubb2008,Barat2000,Berry2015}.
The acceptance angle of our setup for neutral particles is about 70 degrees.
In the off-line data analysis, momentum conservation conditions between detected particles are applied to minimize background signals.
In the measurements, linearly polarized laser pulses (along $z$-direction) are delivered from a home-built Ti:sapphire laser amplifier with a repetition rate of 5 kHz, a central wavelength of 790 nm and a pulse duration of 25 fs.
The laser beam is focused on to the CO$_2$ gas jet in the interaction chamber with a spherical silver mirror which has a focal length of 60 mm.
CO$_2$ molecules are introduced to the interaction chamber from a supersonic gas jet system which consists of a gas nozzle followed by a skimmer. The diameter of the molecular beam is about 170 $\mu$m at the laser focus.
The laser peak intensity at the focus is about $5\times10^{14}$ \Wcm2, so that CO$_2$ molecules can be singly or doubly ionized and subsequently dissociate into fragments. More details about the experimental setup can be found in our previous publications \cite{Eratt2016,Larim2017,Xie2017,Larimian2016,Larimian2017_loc}.


An example for a possible scenario of a LIDR process is depicted in the gray rectangle in Fig. \ref{fig1} (a). After strong-field induced double ionization, CO$_2^{2+}$ can be dissociative and breaks into CO$^+$ and O$^{+}$, which is followed by an electron ({\it e2}) recombining to one of these two ionic fragments to form Rydberg states of O or CO. These processes can be written as CO$_{2}\rightarrow$ CO$^{+}$+O$^{*}$+{\it e} (Ch4) or CO$_{2}\rightarrow$ CO$^{*}$+O$^{+}$+e (Ch5) (see TABLE \ref{tab1} for channel definitions). With the reaction microscope, we achieve complete coincidence detection of the three particles generated in these LIDR processes.

\begin{table}[]
	\centering
	\begin{tabular}{c||l|cc}
		\hline
		\multirow{2}{*}{Channels} & {Dynamics }  & \multicolumn{2}{c}{Branching ratio}\\
		\cline{3-4}
		                       &  {(CO$_{2}\rightarrow$)}   & {Linear}        & {Circular}\\
		\hline
		Ch1                         & CO$^{+}_2$+e                & $24.52\%$    & $46.38\%$ \\
		Ch2                         & CO$^{2+}_2$+2e              & $14.10\%$    & $1.60\%$  \\
		\textbf{Ch3}               & CO$^{+}$+O$^{+}$+2e         & $8.01\%$    & $5.62\%$  \\
		\textbf{\underline{Ch4}} & CO$^{+}$+O$^{*}$+e          & $0.10\%$    & - \\
		\textbf{\underline{Ch5}} & CO$^{*}$+O$^{+}$+e          & $0.002\%$    & - \\
		Ch6                       & CO+O$^{+}$+e                & $16.38\%$    & $12.65\%$ \\
		Ch7                       & CO$^{+}$+O+e                & $36.87\%$    & $33.75\%$ \\
		\hline
	\end{tabular}
	\caption{Fragmentation channels of CO$_2$ in strong laser fields. The branching ratios of each channel in the linear and circular polarization field are shown in the right column. The peak intensity of the circular pulse was 2.5$\times$10$^{14}$W/cm$^2$.}
	\label{tab1}
\end{table}

Figure \ref{fig1}(b) shows the measured photo-ion photo-ion coincidence (PIPICO) distribution. The sharp parabolic PIPICO lines are signals of two-body dissociation channels as indicated in the figure, among which the LIDR channel Ch4 can be clearly identified. The LIDR channel Ch5 is not shown in Fig. \ref{fig1}(b). We will focus on Ch4. Subsequently we will compare it to the other LIDR process Ch5.

No signals of CO$_2$ dissociative recombination in the measurement with circularly polarized laser pulses are observed, as shown the inset of Fig. \ref{fig1}(b). This observation contradicts a previous experimental observation of hydrogen \cite{Zhang2019}, in which Rydberg states are formed via resonant multiphoton excitation by irradiation of the circularly polarized laser fields. Our observation confirms that in our work Rydberg states are formed through frustrated double ionization, which is a rescattering-like process \cite{Nubb2008,Mansch2009,Ulrich2010}. This process will be strongly suppressed for circularly polarized laser fields \cite{HHG1_Corkum1993,Atto1_Corkum2007,Chini2014,Gallmann2012}.


\begin{figure}[tb]
	\centering
	\includegraphics[width=0.9\columnwidth]{{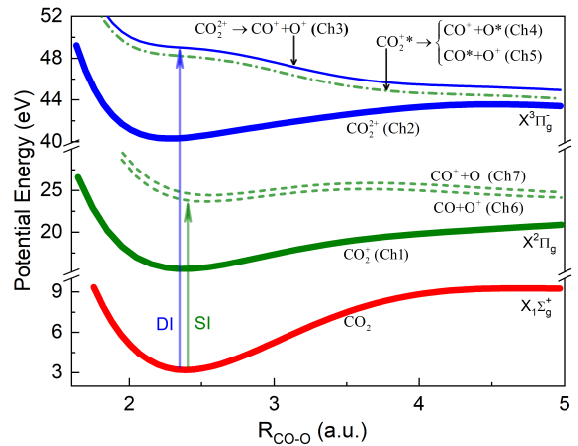}}
	\caption{Simulated potential energy curves of CO$_{2}$, CO$_{2}^+$, and CO$_{2}^{2+}$ over one C-O stretching coordinate. The bold curves are the ground states of CO$_{2}$, CO$_{2}^+$, and CO$_{2}^{2+}$. The dash-dotted curve is a schematic for a high-lying Rydberg state of CO$_{2}^+$ which is directly below the dissociative state of CO$_{2}^{2+}$ (solid blue curve). 
}
	\label{fig2}
\end{figure}

To gain insight into the quantum states involved in our observation of laser induced dissociation of CO$_2$, we performed quantum chemical simulations to obtain the potential energy curves of the relevant states. Multiconfiguration complete-active-space self-consistent field theory (CASSCF) should be employed for an accurate description of the electronic structure of CO$_2$ in excited states. To simulate the dissociation processes of CO$_2$, CO$_2^+$, and CO$_2^{2+}$ along the C-O stretching coordinates in highly-excited states, a multistate complete active space perturbation theory of second order (MS-CASPT2) \cite{finley_multi-state_1998}, which further considered dynamic correlations, was carried out using Molcas 8.2 \cite{aquilante_molcas_2016}. In the active spaces of CASSCF calculations, 10, 9, and 8 electrons in 15 orbitals are included in CO$_2$, CO$_2^+$, and CO$_2^{2+}$ molecules respectively, when they were all kept in C$_2v$ symmetry. In searching for potential energy curves along C-O bonds of CO$_2^+$, the state-averaged CASSCF wave functions for 20 singlet excited states were used in MS-CASPT2 calculations with ANO-RCC-VTZP basis set, while O-C-O valence angles were constrained to 180 degrees. The calculated potential energy curves of CO$_2$, CO$_2^+$, and CO$_2^{2+}$ along the C-O bonds are presented in Fig.~\ref{fig2}.

When a neutral \coo interacts with a strong laser field, single or double ionization may occur through the removing of one or two valence electrons. Since the energy gaps between different MOs are rather small, direct removal of electrons from a low-lying MO is possible, resulting in electronically excited states \cite{Xie2014prl,Xie2014prx}. As shown in Fig.~\ref{fig2}, removal of one or two electrons from the HOMO of \coo leads to stable ground states of the \coo cation or dication, observed as Ch1 or Ch2.
On the other hand, dissociative electronically excited states of cation or dication can be reached through removal of at least one electron from low-lying MOs, yielding Ch3, Ch6 and Ch7.
There is another possible way to populate a molecule in high-lying Rydberg states through the so-called frustrated field ionization.
In case of molecules, such Rydberg states are dissociative and close to the ionization threshold to a higher charge state.
As shown in Fig.~\ref{fig2}, the PEC of a high-lying Rydberg state of \coop is very close to one PEC of an electronically excited state in \coopp. The kinetic energy release (KER) of the dissociation from these two states is similar.


\begin{figure}[tb]
  \centering
  \includegraphics[width=0.9\columnwidth]{{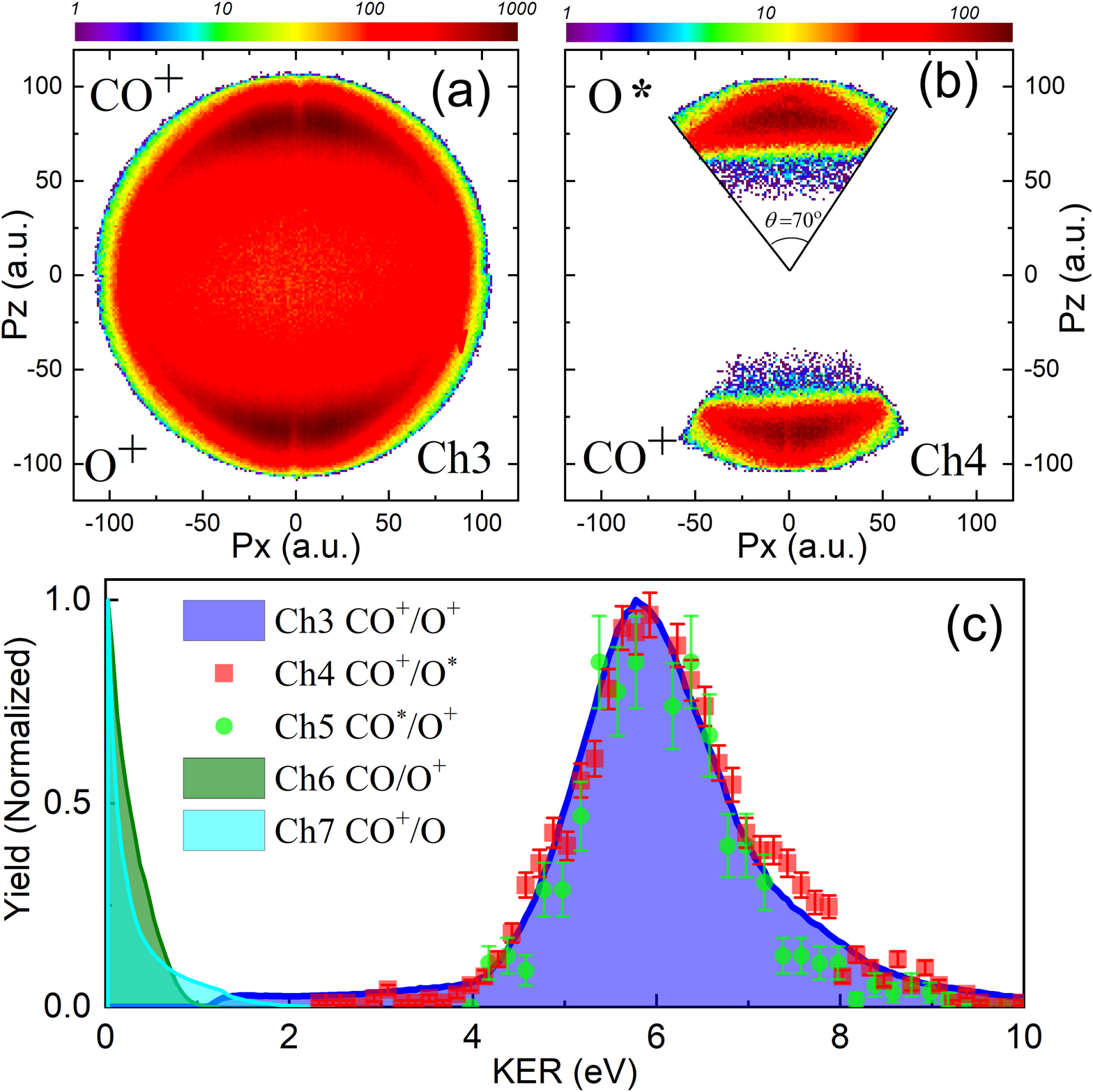}}
  \caption{Measured momentum distributions in the x-z plane for (a) Ch3 and (b) Ch4 with $p_y$ integrated. Laser field polarized along the z-axis direction, indicated by the red arrow. (c) Measured KER spectra of different dissociation channels. The yield of each channel is normalized independently.
    }
  \label{fig3}
\end{figure}

Now we turn to our experimental results.
The momentum distributions of the fragments generated through Ch3 and Ch4 are plotted in Fig.~\ref{fig3} (a-b).
Except the different acceptance angle, the momentum distributions of both channels are similar and peak along the laser-polarized direction.
Such anisotropic distributions indicate electron removal from $\sigma$-type molecular orbitals \cite{Xie2014prl,Eratt2016,Larim2017}.
Since the HOMO of CO$_2$ is a $\pi$-type orbital, these two dissociation channels originate from electronically excited states formed by removal an electron from lower-lying molecular orbitals, which is consistent with the PECs in Fig.~\ref{fig2}.

For further comparison of Ch3 and Ch4, we show the KERs of them in Fig.~\ref{fig3} (c).
One clear observation is that the KER of Ch4 (red square) is almost the same as that of Ch3 (blue area) with the peak at 5.8 eV.
Since the KER distribution is determined by the involved electronic states(see Fig.~\ref{fig2}), this observation indicates that their dissociative nuclear wave packets evolve on PEC with similar shapes.
Furthermore, the same KER distribution of Ch3 and Ch4 not only demonstrates that the recombination of ion and electron occurs during the molecular dissociation, but also provides a clear evidence that the electron is recaptured into such high-lying Rydberg states that the nuclear charge is not fully shielded by the trapped electron leading to similar KER.

On the other hand, the KERs of Ch6 and Ch7 with maxima at zero energy, are much smaller than those of Ch3 and Ch4.
It implies that Ch6 and Ch7 originate from low excited states of CO$^+_2$, as shown in Fig.~\ref{fig2}.
Neutral fragments from those channels are not energetic enough to be detected, therefore only the charged fragments are detected for Ch6 and Ch7 in our experiments.
Moreover, such low-lying electronic excitation can be populated in circular polarized laser fields \cite{Zhang2019}. As shown in Table \ref{tab1}, the branching ratios of Ch6 and Ch7 have no clear dependence on the laser ellipticity. This is very different from the LIDR processes.
The different KER distributions of Ch6 and Ch7 indicate that they originate from different states, which is consistent with Fig.~\ref{fig2}.


\begin{figure}[tb]
  \centering
  \includegraphics[width=0.90\columnwidth]{{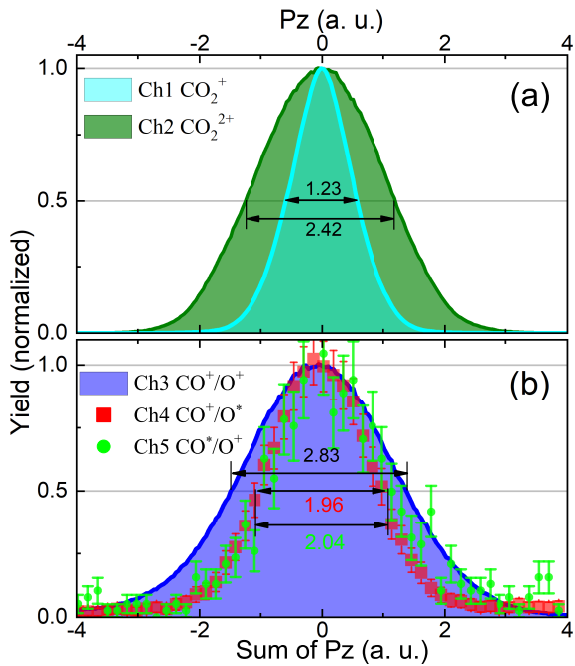}}
  \caption{ Measured ion momentum distributions along the laser polarization direction for Ch1 and Ch2 (a) and ion momentum sum distribution for Ch3, Ch4 and Ch5 (b). The black arrows indicate their full width at half maximum (FWHM). To be noted, the momentum distributions of Ch2 and Ch4 contain the information of the two released electrons.}
  \label{fig4}
\end{figure}

An important and interesting question is which of the released electrons more favorably to recombine with their parent ion?
To answer this question, we analyze the momentum distributions of electrons and ions along the laser polarized direction.
Since the final momentum of photoelectrons from the strong field interaction is determined by the vector potential of the laser field at the birth time of the electron \cite{Goulie2007,Pazourek2015}, electron momentum distributions contain information about the ionization dynamics taking place during the strong field interaction.
The width of the momentum distribution is proportional to $\sqrt{U_p}$, with $U_p$ the ponderomotive potential of the laser pulse.
Due to momentum conservation, the electron momenta can be derived from measured ionic momenta.
Figure \ref{fig4} shows the measured ion momentum distributions of Ch1 to Ch5.
For dissociative processes the momentum sums of the two fragments are employed.
First, we compare the momentum distributions of direct ionization processes (Ch1, Ch2 and Ch3).
Under the used laser conditions, single ionization happens at the leading edge of the laser pulses due to the saturation, which leads to a rather narrow momentum distribution.
On the other hand, in case of double ionization, we notice that the momentum distribution of the dissociative process (Ch3) is broader than that of the non-dissociative process (Ch2).
At the laser intensity of 5$\times$10$^{14}$W/cm$^2$, the dominant double ionization process happens sequentially, which means the two electron are removed one after the other with no correlation \cite{Pei2010}.
That the width of the sum momentum distribution for double ionization is broader than that for single ionization is due to the broader momentum width of the second electron which is removed at a higher laser intensity.
The dissociative process (Ch3) originates from electronically excited states of \coopp, which involves removal of electrons from lower-lying molecular orbitals in the first ionization step with higher ionization potential than removing a HOMO electron \cite{Eratt2016}.
Therefore, dissociative double ionization happens at an effectively higher laser intensity which leads to a broader momentum distribution than that of non-dissociative double ionization.

As already discussed, the dissociative recombination process (Ch4) has the same origin as the dissociative double ionization (Ch3).
Now, we compare their momentum distributions.
It is clear that the momentum distribution of Ch4 is much narrower than that of Ch3.
For Ch4 only one electron is released, with the recombination of either the first or the second ionized electron.
For the double ionization process, the first electron has a narrow momentum distribution (as Ch1) and the second electron has a much broader distribution which leads to a broader momentum sum distribution for double ionization (Ch2 and Ch3).

In the following, we will quantify the contributions of the first and second electrons to the recombination from the measured momentum widths.
Since double ionization happens mainly sequentially in our experiment, it is reasonable to assume that the first ionization step is saturated. 
We now can use the momentum distribution for single ionization (Ch1) as that of the first ionization step ($w_{e1}$).
Then the momentum width of the second ionization step ($w_{e2}$) can be derived from the sum of measured electron momentum width ($w_{e1+e2}$) with the relation of $w_{e1+e2}=\sqrt{w_{e1}^2+w_{e2}^2}$.
From the measured width of the sum momentum distribution of Ch1 (1.24 a.u.) and Ch2 (2.42 a.u.), we obtain the momentum width of the second ionization step of Ch2 to be 2.08 a.u..
With the knowledge that the second ionization of Ch3 is similar to non-dissociative double ionization (Ch2) \cite{Eratt2016}, we used the obtained momentum width of the second ionization step of Ch2 to get the momentum width of the first ionization step of removing low-lying molecular orbitals, which yields 1.92 a.u..
Since only one of these two electrons recombines during dissociative double ionization, the measured electron momentum distribution is determined by the recombination probabilities of the two electrons ($S_{Ch4} = \alpha_{e1} exp(-4\log2p_z^2/w_{e1}^2)+\alpha_{e2} exp(-4\log2p_z^2/w_{e2}^2)$). With the obtained widths of $w_{e1}=1.92\, a.u.$ and
$w_{e2}=2.08\, a.u.$, we performed a fitting of the electron momentum distribution with $S_{Ch4}$ and got the recombination probabilities of the first and second electron: $\alpha_{e1}=77\%$ and $\alpha_{e2}=23\%$. The result shows that the recombination probability of the first electron is about three times as high as that of the second electron.
This is in accord with the electron recapture mechanism, in which only electrons with near-zero momentum can be recaptured.
For the first electron, the momentum distribution has a much narrower width which means the recombination probability of electrons with near-zero momentum is much higher than that of the second electron.
Further pump-probe measurements would be required to gain more insight into the involved electron dynamics.

In the end, we compare two LIDR processes (Ch4 and Ch5).
One interesting observation is that the yield of Ch4 is much higher than that of Ch5, with a yield ratio of about 50 between electron localized to O$^+$ and CO$^+$, as shown in Tab.~\ref{tab1}.
A straightforward explanation of this phenomenon can be that the electronegativity of O$^+$ is higher than that of CO$^+$ which leads to a higher electron recombination probability to O$^+$.
This fact is qualitatively consistent with previous results on low energy electron scattering with \coopp, which shows that the O/CO$^+$ process dominates the dissociative recombination \cite{Seiersen2003}.
This observation implies that the difference in electron recombination probability can be exploited for controlling molecular reactions through laser-induced dissociative recombination.
Except for the much lower yield, Ch5 has a similar KER and a sum momentum distribution as Ch4, as depicted in Fig.~\ref{fig3} and ~\ref{fig4}, respectively, which indicates that the two LIDR processes have a similar laser-induced origin.


In summary, we experimentally investigate the laser-induced dissociative recombination processes of \coo with coincidence detection of all involved particles. Our results provide clear evidences that these processes happen after strong field double ionization followed by the electron recombination. Measurements also show that the recombined electron has much higher probability to become localized on the oxygen site. Analysis of electron momentum distributions show that the first emitted electron has a much higher probability to recombine during the laser-induced dissociative processes. Our work may open a new way of controlling molecular dissociation through electron recombination.

This work was financed by the Austrian Science Fund (FWF) under P25615-N27, P30465-N27, P28475-N27. J. W. thanks the Institute of Organic Chemistry and Biochemistry, Academy of Sciences of the Czech Republic (RVO: 61388963).


\end{document}